\documentclass[8pt]{elsart}
\usepackage{graphicx}
\usepackage{longtable}
\usepackage{amsmath,amssymb}
\usepackage{dcolumn}
\usepackage{cite}
\usepackage{amsfonts}
\usepackage{hangcaption}
\usepackage[english]{babel}
\usepackage[latin1]{inputenc}

\newcommand{\be}{\begin{eqnarray}}
\newcommand{\ee}{\end{eqnarray}}

\newcommand{\mat}{\left ( \begin{array}{cc}}
\newcommand{\emat}{\end{array} \right )}
\newcommand{\vect}{\left ( \begin{array}{c}}
\newcommand{\evect}{\end{array} \right )}

\def\PL{{\it Phys. Lett. }}
\def\NP{{\it Nucl. Phys. }}

\begin{document}

\begin{frontmatter}
\title{The Quantum Viscosity Bound In Lovelock Gravity}

\author[cqupt,email1]{Fu-Wen Shu}
\address[cqupt]{College of Mathematics and Physics,Chongqing University of
Posts and Telecommunications, Chongqing, 400065, China}
\thanks[email1]{e-mail: shufw@cqupt.edu.cn}

\begin{abstract}
Based on the finite-temperature AdS/CFT correspondence, we calculate
the ratio of shear viscosity to entropy density in any Lovelock
theories to any order. Our result shows that any Lovelock correction
terms except the Gauss-Bonnet term have no contribution to the value
of $\eta/s$. This result is consistent with that of Brustein and
Medved's prediction.

\noindent PACS: number(s): 11.25.Tq; 04.50.-h; 04.70.Dy; 11.25.Hf

Keywords: AdS/CFT, KSS bound, Lovelock theory, AdS black brane.
\end{abstract}
\end{frontmatter}

\maketitle

Stimulated by the conjecture of AdS/CFT correspondence
\cite{ads/cft,gkp,w}, string theory has attracted a lot of
attention, especially after the discovery that some theoretical
results of the dual theory are consistent with that of the RHIC
experiment, say, the ratio of the viscosity to the entropy density
\cite{pss0,kss}. Recently, it was conjectured, based on the AdS/CFT
correspondence, that for all possible nonrelativistic fluids, there
may exist a universal lower bound (the KSS bound) on the
viscosity/entropy-density ratio (we set
$G=c=\hbar=k_B=1$)\cite{kovtun}
\begin{equation}
\frac{\eta}{s}=\frac{1}{4\pi}.
\end{equation}
This bound received great supports from several kinds of field
theories \cite{cch,fbb,myers0}, as well as the case with chemical
potential in the theory\cite{gmsst,myers1}. However, more recent
work on the higher derivative gravity theories (see
\cite{kp,shenker,shenker1,gmsst1,gs1,myers2,ohta}) showed that the
KSS bound is violated when the dual gravity is enlarged to include a
stringy correction (see \cite{dutta} for more about the KSS bound in
higher derivative gravity). This correction is frequently referred
to as the quantum correction, since in CFT side this is a correction
of the 't Hooft coupling $\lambda=g_{YM}^2N_c$. It is of particular
significance to consider the $1/\lambda$ correction when we are
dealing with non-extremely strong coupling fluids.

Recently, the authors of \cite{brustein} predicted that all Lovelock
terms higher than the second order(the Gauss-Bonnet term) do NOT
contribute to the value of $\eta/s$ at all, and this prediction was
partially confirmed in \cite{gs2} for the third-order Lovelock
gravity. In this paper we calculate the viscosity/entropy-density
ratio directly in the Lovelock theory to any order, trying to make a
complete verification of the prediction, and indeed, our result
provides a direct support of this prediction as will see below.

We start with the Lovelock theory of gravity. This is one of the
most general second order gravity theories in higher dimensional
spacetimes and is free of ghost when expanding on a flat
space\cite{zwiebach} and hence is of particular interest. The
Lagrangian density for general Lovelock gravity in $D$ dimensions is
${\mathcal{L}}=\sum_{m=0}^{[D/2]}c_{m}\,{\mathcal{L}}_{m},$ where
${\mathcal{L}}_{m}$ is given by \cite{lovelock}
\begin{equation}\label{LUVLagrangian}
{\mathcal{L}}_{m}=\frac{1}{2^m}\sqrt{-g}\delta^{\lambda_1 \sigma_1
\cdots \lambda_m \sigma_m}_{\rho_1 \kappa_1 \cdots \rho_m \kappa_m}
       R_{\lambda_1 \sigma_1}{}^{\rho_1 \kappa_1} \cdots  R_{\lambda_m \sigma_m}{}^{\rho_m \kappa_m}\,,
\end{equation}
$c_{m}$ is the $m$'th order coupling constant, $[D/2]$ denotes the
integer value of $D/2$ and the Greek indices $\lambda$, $\rho$,
$\sigma$ and $\kappa$ go from $0$ to $D-1$. The symbol $R_{\lambda
\sigma}{}^{\rho \kappa}$ is the Riemann tensor in $D$-dimensions and
$\delta^{\lambda_1 \sigma_1 \cdots \lambda_m \sigma_m}_{\rho_1
\kappa_1 \cdots \rho_m \kappa_m}$ is the generalized totally
antisymmetric Kronecker delta. The term
${\mathcal{L}}_{0}=\sqrt{-g}$ is the cosmological term, while
${\mathcal{L}}_{1}=\sqrt{-g}\delta_{\rho_1\kappa_1}^{\lambda_1\sigma_1}\,R_{\lambda_1\sigma_1}{}^{\rho_1\kappa_1}/2$
is the Einstein term. In general ${\mathcal{L}}_m$ is the Euler
class of a $2m$ dimensional manifold.

Variation of the Lagrangian with respect to the metric yields the
Lovelock equation of motion
\begin{eqnarray}
    0={\mathcal{G}}_{\mu}^{\nu}
      =-\sum_{m=0}^{[D/2]}\frac{c_m}{2^{(m+1)}}
     \delta^{\nu \lambda_1 \sigma_1 \cdots \lambda_m \sigma_m}_{\mu \rho_1 \kappa_1 \cdots \rho_m \kappa_m}
       R_{\lambda_1 \sigma_1}{}^{\rho_1 \kappa_1} \cdots  R_{\lambda_m \sigma_m}{}^{\rho_m \kappa_m}  \ , \label{eq:EOM}
\end{eqnarray}

As is shown in \cite{Wheeler}, there exist static exact solutions of
Lovelock equation. Let us consider the following metric
\begin{eqnarray}
   ds^2=-f(r)dt^2 + \frac{dr^2}{f(r)}+r^2\sum_{i,j}^{D-2}\gamma_{ij}dx^idx^j, \label{eq:solution}
\end{eqnarray}
where $\gamma_{ij}dx^idx^j$ represents the line element of a
$(D-2)$-dimensional Einstein space. With this ansatz, we have
\begin{equation}
\label{ricci einstein}
\mathcal{R}_{ijkl}=\kappa(\gamma_{ik}\gamma_{jl}-\gamma_{il}\gamma_{jk}),\
\ \mathcal{R}_{ij}=\kappa(D-3)\gamma_{ij},\ \
\mathcal{R}=\kappa(D-2)(D-3).
\end{equation}
where $\kappa$ is the curvature constant, whose value determines the
geometry of the horizon. Without loss of the generality, one may
take $\kappa=1$, $0$, or $-1$ representing sphere, flat and
hyperbolic respectively.

Using this metric ansatz, we can calculate Riemann tensor components
as
\begin{eqnarray}
    R_{tr}{}^{tr}=-\frac{f^{''}}{2},\ R_{ti}{}^{tj}=R_{ri}{}^{rj}=-\frac{f^{'}}{2r}\delta_{i}{}^{j},\ R_{ij}{}^{kl}=\left(\frac{\kappa-f}{r^2}\right)\left(\delta_{i}{}^{k}\delta_{j}{}^{l}-\delta_{i}{}^{l}\delta_{j}{}^{k}\right) \ . \label{eq:riemann}
\end{eqnarray}
Substituting (\ref{eq:riemann}) into (\ref{eq:EOM}) derives a simple
equation
\begin{eqnarray}\label{eom}
W[\psi]\equiv \sum_{m=0}^n\tilde{c}_m\psi^m=\frac{\mu}{r^{D-1}},
\end{eqnarray}
where $\psi=r^{-2}(\kappa-f)$, $\mu>0$ is a constant of integration
 which is related to the ADM mass by
\begin{eqnarray}
    M=\frac{\mu V_{D-2}}{16\pi G_D} \ , \label{eq:ADM}
\end{eqnarray}
where $V_{D-2}$ is the volume of the $(D-2)$-dimensional
hypersurface and $G_D$ is the Newton constant. In (\ref{eom}), we
also defined $\tilde{c}_m\equiv \frac{(D-3)!}{(D-2m-1)!}c_m$ and $n$
is an integer with $0<n\leq [D/2]$. In this paper we are considering
AdS black brane in Lovelock gravity, so we have $c_0=-2\Lambda$ with
the cosmological constant $\Lambda=-(D-1)(D-2)/2l^2$ and $c_1=1$.

 We would like to extract some information from the Lovelock black
brane, such as their thermodynamic properties. One quantity which is
of particular interest is the entropy $S$. Generally speaking, one
can obtain the entropy of a black hole in higher derivative theories
by using the thermodynamic relation $S=-\partial F/\partial T$ with
$F$ the free energy and $T$ the Hawking temperature. By doing so one
finds that the entropy of the Lovelock black brane is given
by\cite{cai}
\begin{eqnarray}\label{entropy}
S=\frac{V_{D-2}r_+^{D-2}}{4G_D}\sum_{m=1}^{n}\frac{m(D-2)}{(D-2m)}
\tilde{c}_m(\kappa r_+^{-2})^{m-1},
\end{eqnarray}
where $r_+$ is the event horizon of the black brane which is the
positive root of $f(r_+)=0$. In the present paper, we mainly focus
on the case where $\kappa=0$. In this case we have a simple formula
for the entropy density of the Lovelock black brane
\begin{eqnarray}\label{entropyden}
s=\frac{r_+^{D-2}}{4G_D},
\end{eqnarray}
and now, $r_+$ is a solution of $\psi(r_+)=0$. The Hawking
temperature of this case is given by
\begin{eqnarray}\label{temp}
T=\frac{(D-1)\tilde{c}_0}{4\pi}r_+.
\end{eqnarray}
In what following, we would like to see the waves generated by a
metric perturbation of the background. Generally speaking, there are
scalar, vector and tensor modes depending on the rotation symmetry.
In this paper, we only study tensor perturbations which is closely
related to the shear viscosity as will see below.

We now add a small tensor perturbations to the solution
(\ref{eq:solution})
\begin{eqnarray}
 \delta g_{i j}=r^2 \phi(t,r)h_{i j}(x^i ) \ ,\ \ \  (others)=0
\end{eqnarray}
where $\phi (t,r)$ represents the dynamical degrees of freedom.
Here, $h_{ij}$  are defined by
\begin{eqnarray}
   \nabla^{k}\nabla_{k}h_{ij}=k^2 h_{ij} \ , \qquad
    \nabla^{i} h_{ij}=0 \ ,\quad \gamma^{ij}h_{ij}=0.
\end{eqnarray}
Here, $\nabla^{i}$ denotes a covariant derivative with respect to
 $\gamma_{ij}$ and $k^2$ is the eigenvalue playing a role of momentum.

With these definition, one can obtain the first order perturbation
equation of the Lovelock equation (\ref{eq:EOM})\cite{soda}
\begin{eqnarray}
    0=\delta{\mathcal{G}}_{\mu}^{\nu}=-\sum_{m=1}^{k}\frac{a_m}{2^{(m+1)}}
     \delta^{\nu \lambda_1 \sigma_1 \cdots \lambda_m \sigma_m}_{\mu \rho_1 \kappa_1 \cdots \rho_m \kappa_m}
       R_{\lambda_1 \sigma_1}{}^{\rho_1 \kappa_1} \cdots  R_{\lambda_{m-1} \sigma_{m-1}}{}^{\rho_{m-1} \kappa_{m-1}} \delta R_{\lambda_m \sigma_m}{}^{\rho_m \kappa_m},  \label{eq:pert}
\end{eqnarray}
where $\delta R_{a b}{}^{c d}$ represents the first order variation
of the Riemann tensor and we have introduced a new quantity
$a_m=mc_m(m>0)$. As shown in \cite{soda}, it is straightforward once
we know the expressions of quantities
 $\delta R_{ti}{}^{tj}$, $\delta R_{ri}{}^{rj}$ and $\delta
 R_{ij}{}^{kl}$. Then the calculation becomes a mathematical
 game and the result is ready-made\cite{soda}
\begin{eqnarray}0=\delta{\mathcal{G}}_{i}{}^{j}=\frac{1}{r^{D-4}}\left[\frac{h}{2f}\left(\ddot
\phi-f^2\phi^{''}\right)-\left(\frac{(r^2fh)'}{2r^2}\right)\phi^{'}+\frac{(k^2+2\kappa)h^{'}}{2(D-4)r}\phi\right]h_{i}{}^{j},\label{eq:deltaG}
\end{eqnarray}
where
\begin{eqnarray}
\nonumber  h(r) &=& \frac{d}{dr}\left[ \frac{r^{D-3}}{D-3}  \frac{dW[\psi]}{d\psi} \right] \\
  &=&r^{D-4}-\sum_{m=2}^n\Biggl[\frac{m\tilde{c}_mr^{D-2m-2}(\kappa-f)^{m-2}}{D-3}
\left\{(m-1)rf^{'}-(D-2m-1)(\kappa-f)\right\}\Biggr].
  \label{h:W}
\end{eqnarray}
Using the Fourier decomposition
\begin{eqnarray}
\phi(t,r)=\int\frac{d\omega}{2\pi}e^{-i\omega t}\phi(r),
\end{eqnarray}
we obtain the linearized equation of motion for $\phi(r)$:
\begin{eqnarray}
  \phi''(r)
      + \left( \frac{(r^2fh)^{'}}{r^2fh} \right) \phi'(r)
    +\frac{1}{f^2}\left(\omega^2-\frac{(k^2+2\kappa)fh^{'}}{(D-4)rh}\right) \phi(r)=0 \ .
 \label{eq:master_eq}
\end{eqnarray}
It is convenient to introduce a new dimensionless coordinate
$u=(r_+/r)^{(D-1)/2}$ with $r_+$ the event horizon of the black
brane. In this coordinate frame, $u=0$ corresponds to the boundary
and $u=1$ the horizon. The linearized equation of motion
(\ref{eq:master_eq}) then becomes (for $\kappa=0$)
\begin{eqnarray}
   \phi''(u)
      + \frac{g'(u)}{g(u)}  \phi'(u)
    +\frac{\bar{\omega}^2}{u^{\frac{2D-6}{D-1}}\psi^2(u)}\phi(u)-\frac{D-1}{2(D-4)}\cdot
\frac{h'\bar{k}^2}{u^{\frac{D-5}{D-1}}\psi(u) h} \phi(u)=0
    \,
 \label{eom1}
\end{eqnarray}
where
\begin{eqnarray}
&&\label{gfun} g(u)=-r_+^{4-D}\psi(u)h(u)u^{\frac{D-7}{D-1}},\\
&&\bar{\omega}\equiv \frac{2}{(D-1)r_+}\omega,\ \ \ \ \bar{k}\equiv
\frac{2}{(D-1)r_+}k,
\end{eqnarray}
and the prime denotes the derivative with respect to $u$.

Now we shall calculate the shear viscosity in Lovelock gravity
theories. Generally speaking, the shear viscosity $\eta$ can be
calculated via Kubo formula,
\begin{equation}
\eta=-\lim_{\omega\rightarrow 0} \frac{\mbox{Im}(G^R(\omega,
0))}{\omega}, \label{kubo}
\end{equation}
where $G^R$ is the retarded Green's function
\begin{equation}
G^R(\omega,\vec{k})=-i\int
dtd\vec{x}e^{-i\vec{k}.\vec{x}}\theta(t)<[\hat{\mathcal{O}}(x)\hat{\mathcal{O}}(0)]>,
\end{equation}
with $\hat{\mathcal{O}}$ some boundary CFT operators. According to
AdS/CFT correspondence, the Green's function can be calculated from
the dual gravity side via the Gubser-Klebanov-Polyakov/Witten
relation~\cite{gkp, w}
$$
\langle e^{\int_{\partial M}\phi_0
\hat{\mathcal{O}}}\rangle=e^{-S_{cl}[\phi_0]},
$$
where $\phi$ is the bulk field and $\phi_0$ is its value at the
boundary, i.e., $\phi_0=\lim_{u\rightarrow0}\phi(u)$. Extracting the
part of $S_{cl}$ that is quadratic in $\phi$ and inserting the
solution of the linearized field equation we may get a surface term
in four dimensions by using the equation of motion,
\begin{equation}
S_{cl}[\phi_{0}] =\!\int\!\frac{ d^{D-1}k}{(2\pi)^{D-1}}\phi_{0}(-k)
G(k,u)\phi_{0}(k) \bigg|_{u=0}^{u=1},
\end{equation}
where $ u=(r_+/r)^{(D-1)/2}$ as defined previously.
In this way, we obtain the following relation for the retarded
Green's function\cite{ss}
\begin{equation}
G^R(k)=2G(k, u)\bigg|_{u=0}, \label{green}
\end{equation}
where the incoming boundary condition at the horizon is imposed. The
shear viscosity then can be calculated by using (\ref{kubo}).

In the following we would like to calculate the shear viscosity,
following the procedures introduced above. The main task is to solve
the equation of motion (\ref{eom1}) in hydrodynamic regime
$\it{i.e.}$, \ small $\omega$ and $k$. To solve the wave equation
(\ref{eom1}) we first examine the behavior around the horizon where
$u=1$. For this purpose it is convenient to impose a solution as
\begin{equation}\label{solution}
\phi(u)=(1-u)^\nu F(u),
\end{equation}
with $F(u)$ regular at the horizon. Substituting (\ref{solution})
into the wave equation (\ref{eom1}) and leaving the most divergent
terms, we can obtain
\begin{eqnarray}\label{nu1}
 \nu=\pm i\frac{\bar{\omega}}{\psi'(1)}
    \,
 \label{nu}
\end{eqnarray}
where we have used the relations
\begin{eqnarray}\label{g1}g(u\rightarrow
1)&=&-g'(1)(1-u)+\mathcal{O}((1-u)^2),\\
\label{f1}\psi(u\rightarrow
1)&=&-\psi'(1)(1-u)+\mathcal{O}((1-u)^2).
\end{eqnarray}
In present paper we choose $``-''$ sign in eq. (\ref{nu1}) for
convenience.

To get the viscosity via Kubo formula (\ref{kubo}), the standard
procedure is to consider series expansion of the solution in terms
of frequencies up to the linear order of $\omega$,
\begin{equation}
F(u) =F_0(u)+\nu F_1(u) + {\mathcal{O}}(\nu^2, k^2). \label{series}
\end{equation}
Then the equation of motion (\ref{eom1}) becomes the following form
up to ${\mathcal{O}}(\nu)$, 
\begin{equation}
\label{eq} \left[g(u)F'(u)\right]'
-\nu\left(\frac{1}{1-u}g(u)\right)'F(u)-\frac{2\nu}{1-u}g(u)F'(u)=0.
\end{equation}
After substituting the series expansion (\ref{series}) into the
equation (\ref{eq}), we obtain the following equations of motion for
$F_0(u)$ and $F_1(u)$
\begin{eqnarray}
\left[g(u)F'_0 (u)\right]'=0,\\
\left[g(u)F'_1 (u)\right]'-\left(\frac{1}{1-u}g(u)\right)'F_0(u)=0.
\end{eqnarray}
By requiring that the functions $F_0 (u)$ and $F_1(u)$ are regular
at the horizon one gets the following results
\begin{eqnarray}
F_0 (u)=C,\\
F_1' (u)=\left(\frac1{1-u}+\frac{g'(1)}{g(u)}\right)C,
\end{eqnarray}
where again we have used the relation (\ref{g1}) and the constant
$C$ can be determined in terms of boundary value of the field, i.e.,
$
C=\phi_{0}\Big(1+{\mathcal{O}}(\nu)\Big).
$

Now we shall calculate the retarded Green's function. Using the
equation of motion, the action reduces to the surface terms. The
relevant part is given by
\begin{equation}
S_{cl}[\phi(u)]= -\frac{(D-1)r^{D-1}_{+}}{64\pi G_D}
\!\int\!\frac{d^{D-1} k}{(2\pi)^{D-1}}
\Big(g(u)\phi(u)\phi'(u)+\cdots\Big)\Bigg|_{u=0}^{u=1}.
\end{equation}
Near the boundary $u=\varepsilon$, using the perturbative solution
of $\phi(u)$, we get
\begin{eqnarray}
\phi'(\varepsilon) &=& \nu\frac{g'(1)}{g({\varepsilon})}\phi_{0}
+{\mathcal{O}}(\nu^2, k^2) \nonumber
\\
&=& -i\frac{\bar{\omega}}{\psi'(1)}
\frac{g'(1)}{g(\varepsilon)}\phi_{0} +{\mathcal{O}}(\omega^2, k^2).
\end{eqnarray}
Therefore we can read off the correlation function from the relation
(\ref{green}),
\begin{equation}
G^R(\omega, k) =i\omega\frac{1}{16\pi
G_D}\left(\frac{r_+^{D-2}}{\psi'(1)}\right)g'(1) +{\mathcal{
O}}(\omega^2, k^2),
\end{equation}
where contact terms are subtracted. Then the shear viscosity can be
obtained by using Kubo formula (\ref{kubo}),
\begin{equation}\label{viscosity}
\eta=-\frac{1}{16 \pi
G_D}\left(\frac{r^{D-2}_{+}}{\psi'(1)}\right)g'(1).
\end{equation}
The ratio of the shear viscosity to the entropy density is concluded
as
\begin{equation}
\frac{\eta}{s} =-\frac{1}{4 \pi }\frac{g'(1)}{\psi'(1)}.
\end{equation}
From (\ref{gfun}) we have a relation $ g'(1)=-r_+^{4-D}\psi'(1)h(1)$
and $h(1)$ can be obtained from (\ref{h:W}) by inserting $\kappa=0$
$$
h(1)=r_+^{D-4}\Big(1-(D-1)(D-4)\tilde{c}_0a_2\Big).
$$
It is straightforward to show that
\begin{equation}\label{final}
\frac{\eta}{s} =\frac{1}{4 \pi
}\Big(1-(D-1)(D-4)\tilde{c}_0a_2\Big)=\frac{1}{4 \pi
}\Big(1-\frac{2(D-1)(D-4)\lambda}{l^2}\Big),
\end{equation}
where we have defined $\lambda=c_2$. This result is exactly the one
predicted in \cite{brustein}.

In summary, we have computed the ratio of shear viscosity to entropy
density for any Lovelock theories. Our result shows that any
correction terms except the Gauss-Bonnet term do not affect the
value of $\eta/s$, and this confirms the prediction made by
\cite{brustein}. During our calculation, we have chosen a vanishing
curvature constant $\kappa$. Actually, our result is still valid
(for leading term) for nonzero $\kappa$ if we focus on a large black
brane. In the large black brane limit, both the entropy and the
viscosity have the same leading terms as those of $\kappa=0$. This
can be seen by noting the expressions of entropy density and
viscosity. From \eqref{entropy}, the entropy density of the Lovelock
black brane with nonzero $\kappa$ can be expanded, in the large
black brane limit($\it{i.e.}$, $\frac{\kappa}{r_+^2}\ll 1$), to the
first order as
\begin{equation}\label{sk}
s=\frac{r_+^{D-2}}{4G_D}\left[1+\frac{2(D-2)\tilde{c}_2}{D-4}\cdot\frac{\kappa}{r_+^2}\right]+\mathcal{O}\left(\frac{\kappa}{r_+^2}\right).
\end{equation}
With the same spirit one can also expand the shear viscosity to the
first order in the large black brane limit. This can be done by
repeating the previous procedures and noting that $h(1)=h(u=1)$ can
be obtained from \eqref{h:W}. In this way, the shear viscosity for
nonvanishing $\kappa$ can be expanded to the first order as
\begin{eqnarray}\label{ek}
\nonumber\eta&=&\frac{r_+^{D-2}}{16\pi
G_D}\left\{1-\frac{2(D-1)(D-4)\lambda}{l^2}-\right.\\
&&\left.\frac{2(D-1)}{D-3}\left[\tilde{c}_2(1-2\tilde{c}_2)+3\frac{\tilde{c}_3}{l^2}-(D-5)\tilde{c}_2\right]\cdot\frac{\kappa}{r_+^2}\right\}+\mathcal{O}\left(\frac{\kappa}{r_+^2}\right).
\end{eqnarray}
From \eqref{sk} and \eqref{ek} it is obvious that the leading terms
of the entropy density and the viscosity for $\kappa\neq 0$ are the
same as those of $\kappa=0$. In other words, in the large black
brane limit, the curvature constant $\kappa$ has no contribution to
the shear viscosity to entropy density ratio for the leading term.
The sub-leading terms, however, receive contributions from $\kappa$.

So far we are confident with the violation of the KSS bound while we
are not sure the existence of a universal lower bound of $\eta/s$. A
great progress alone this line appeared several months ago when the
authors of \cite{brustein1} gave a proof for the existence of a
universal bound of $\eta/s$ for any ghost-free extension of Einstein
theory. However, the work in\cite{shenker1} shows that the causality
violation of the dual gauge theory may put constraints on the
coefficients of higher derivative terms and this in turn will put
constraints on the value of $\eta/s$. Then it is natural to ask if
the lower bound still exists as these constraints are taken into
account. Recent progress made by Camanho and Edelstein in \cite{CE}
provides us with an answer that the causality violation, as
expected, may impose a constraint on the bound of the $\eta/s$ at
least for cubic Lovelock gravity. For completeness, we briefly catch
some important results from \cite{CE}, so as to compare the lower
limit on $\eta/ s$ from causality violation and the result in the
present paper. Actually, the formula of the ratio between $\eta$ and
$s$ obtained in \cite{CE} is not different from our result
\eqref{final}. What is new of their result is that by imposing a
condition so that the causality violation can be avoided, they found
constraints on the coefficient $\lambda$ (or $c_2$ as defined) in
\eqref{final}. For any order ($n\geq 2$) Lovelock gravity, the
condition to be free of causality is that
\begin{equation}\label{causality}
\sum_{m=1}^{n}mc_m\Lambda^{m-1}\left(1+\frac{\gamma(m-1)(D-1)}{D-3}\right)\geq0,
\end{equation}
where $\gamma=-2,-1,2/(D-4)$ represent helicity zero, helicity one
and helicity two graviton, respectively. Therefore, though any
correction terms higher than the second order of Lovelock gravity do
not manifestly contribute to the ratio of viscosity to entropy
density, it does not mean that they are irrelevant to this ratio.
Through \eqref{causality} we see these terms impose a constraint on
the value of $c_2$ (or $\lambda$) thus in turn affecting the lower
bound for $\eta/s$.


\begin{center}\textbf{ACKNOWLEDGEMENTS}\end{center}
\hspace*{7.5mm} The author would like to thank Profs. Y.-G. Gong and
S.-J. Sin for their valuable comments. This work was supported in
part by Natural Science Foundation Project of CQ CSTC under Grant
No. 2009BB4084 and key project from NNSFC (No. 10935013).

\end{document}